\documentclass[10pt,aps,prc,twocolumn,amsmath,amssymb,floatfix,superscriptaddress]{revtex4-1}
\usepackage{mathptmx}
\usepackage{CJK}
\usepackage{graphicx}
\usepackage{bm}
\usepackage{booktabs,array}
\usepackage[breaklinks,pdfstartview=FitH,bookmarksnumbered=true,
bookmarksopen=true,colorlinks=true,linkcolor=blue,urlcolor=blue,citecolor=blue]{hyperref}
\DeclareMathAlphabet{\mathcal}{OMS}{cmsy}{m}{n}
\newcommand{\Ti}{\ensuremath{^{46}\mathrm{Ti}}}
\newcommand{\Ca}{\ensuremath{^{42}\mathrm{Ca}}}
\newcommand{\dd}{\mathrm{d}}
\newcommand{\kap}{\kappa_s}
\begin{document}

\title{Spatial asymmetry of neutron pairing:A distance-resolved study of $^{46}\mathrm{Ti}$}

\begin{CJK*}{UTF8}{gbsn}
\author{Huai-Tong~Xue~(薛怀通)}\email[Contact author:~]{htxue@nynu.edu.cn}
\author{Hua-Ming~Dai~(戴华名)}
\author{Wan~Chang~(常婉)}
\author{Zi-Le~Zhang~(张紫乐)}
\author{Shuai-Wei~Wang(王帅伟) }\email[Contact author:~]{shuaiweiwang@sina.com}
\affiliation{College of Physics and Electronic Engineering, Nanyang Normal University, Nanyang 473061, China}
\begin{abstract}
Distance integration can conceal the spatial organization of neutron pairing in a reflection-asymmetric intrinsic configuration. We resolve the spin-singlet pairing tensor by relative distance and signed pair-center position along a prescribed $\alpha+\Ca$-like path in $\Ti$, using density-constrained Skyrme Hartree--Fock plus Bardeen--Cooper--Schrieffer calculations. Compact and extended components have oppositely signed shifts relative to the normal-density product conditioned on the same distance interval. At a reference center separation of 6 fm, the distance-integrated difference is 0.010, whereas the compact--extended contrast is 0.439. For the baseline surface-pairing prescription, this contrast increases between reference center separations of 6 and 7 fm while the extended anomalous weight decreases and the average neutron gap increases. Endpoint partitions locate an excess of extended neutron weight with both nucleons on the core side. An interference-retaining analysis in the stored canonical representation links the anomalous redistribution to $f_{7/2}$-weighted branches gaining core-side support with distinct amplitude responses. Changes in the normal-density reference contribute separately to the growing contrast. The conditional neutron pattern persists under the tested functional and pairing variations. Distance-conditioned distributions thus reveal intrinsic pairing structure hidden by scalar measures and distance integration.
\end{abstract}
\maketitle
\end{CJK*}

\section{Introduction}
Pairing correlations are a central ingredient of nuclear mean-field descriptions \cite{Dean2003,Bender2003RMP}. Pairing energies and average gaps characterize their strength in open-shell nuclei \cite{Bender2000Pairing}. Their dynamical role is evident in enhanced two-nucleon transfer \cite{Scamps2013} and in microscopic fission calculations with coupled shape and pairing modes \cite{Bulgac2016}. The anomalous density, or pairing tensor, $\kappa(\bm r_1,\bm r_2)$ describes spatial structure through the relative distance of the two nucleons and the position of their center. Coordinate-space calculations have traced compact neutron correlations to the coherent superposition of many single-particle configurations \cite{Matsuo2005}. Studies of low-density matter and halo nuclei connect the pair structure to density and location relative to the core \cite{Matsuo2006,Hagino2007}. Analyses of the nonlocal tensor also reveal pronounced spatial variations in finite nuclei \cite{Pillet2007}. Strength, relative extent, and spatial distribution characterize distinct properties of pairing.

The relation between pair size and nuclear geometry has been investigated in spherical and deformed nuclei \cite{Pillet2007,Pillet2010}. Small pair sizes near the nuclear surface reflect both pairing correlations and the confinement and spatial extent of single-particle wave functions \cite{Pillet2010,Vinas2010}. The local pairing profile can also differ from the profile of the attractive interaction \cite{Sandulescu2005}; bare and induced interactions contribute differently to the spatial pairing field \cite{Pastore2008}. Direction-resolved calculations in deformed nuclei already demonstrate the sensitivity of the nonlocal tensor to the orientation of its two coordinates \cite{Pillet2010}. Radial averaging removes the distinction between unequal ends. Integration over relative distance can further conceal components with different spatial preferences.

The question addressed here is how compact and extended neutron pairing components are allocated between the unequal ends of a reflection-asymmetric intrinsic nucleus. Such preferences could cancel in a distance-integrated measure. We retain the signed axial position of the pair center together with the relative distance. We compare the pair-center allocations of the pairing tensor and the normal-density product within the same relative-distance windows. This comparison distinguishes their conditional spatial distributions from their different relative-distance compositions.

Unequal two-center configurations connect nuclear geometry with microscopic structure \cite{Butler1996,Freer2018}. Density-functional studies describe cluster-like structures associated with expansion and deformation \cite{Girod2013,Ebran2014}. Prescribed-density Hartree--Fock calculations minimize the compound energy at neutron and proton densities assembled from constituent nuclei \cite{Umar2023prc}. Localization functions \cite{Reinhard2011,Zhang2016} and orbital dispersion \cite{Ebran2013prc} characterize the normal-density structure; beyond-mean-field calculations connect intrinsic localization with collective states in $^{12}$C and $^{20}$Ne \cite{Xue2024C12,Xue2025Ne}. In titanium, alpha-core potential studies \cite{Bai2021} and microscopic surface-density calculations \cite{Horiuchi2022} examine how clustering depends on shell structure. The $\alpha+\Ca$ channel supplies both unequal spatial regions and an open neutron shell for studying anomalous correlations. Transfer-reaction angular correlations identify alpha-cluster candidates in $\Ti$ at excitation energies of 11--17 MeV \cite{Fukada2009}. Motivated by this channel, we use translated fragment densities to prescribe a sequence of static $\alpha+\Ca$-like intrinsic configurations. At each prescribed geometry, the compound orbitals and pairing amplitudes adjust self-consistently under the normal-density constraint, allowing a controlled comparison of anomalous correlations across the sequence.

We calculate neutron and proton pairing along this path using a density-constrained axial Skyrme Hartree--Fock (SHF) approach combined
with Bardeen--Cooper--Schrieffer (BCS) pairing. Distance-conditioned distributions resolve how pairing components are allocated between
the unequal ends. Endpoint partitions and interference-retaining orbital budgets connect these allocations to pair geometry and
single-particle structure. Functional comparisons examine the sensitivity of this spatial organization to the underlying
single-particle spectrum and pairing amplitudes.

Section~\ref{Theoretical} defines the model and spatial measures. Section~III presents the intrinsic path, conditional distributions, and microscopic analysis. Section~\ref{Physical} discusses their physical interpretation and scope; Sec.~\ref{Summary} summarizes the findings. The Supplemental Material documents the numerical controls and additional diagnostics.

\section{Theoretical framework}\label{Theoretical}
\subsection{Density-constrained SHF+BCS calculations}
The axially symmetric, time-reversal-invariant SHF+BCS equations are solved for $Z=22$ and $N=24$, with separate neutron and proton pairing. Reflection symmetry is not imposed. The SLy5 energy density functional (EDF) \cite{Chabanat1998} is employed as the baseline particle--hole interaction. Its central quadratic spin-current terms are retained, and no additional tensor force is introduced. The reference density is constructed by translating and summing the independently calculated $^4$He and $\Ca$ densities. Their centers satisfy $4z_\alpha+42z_{\rm Ca}=0$, with the alpha particle on the negative side, and
\begin{equation} d=z_{\rm Ca}-z_\alpha. \end{equation}
Here $d$ is the center-to-center separation of the reference fragments, defined before the compound density is optimized. It differs from their surface separation. Shape-preserving cubic interpolation translates the fragment densities, with zero density outside their original support. Their particle numbers are checked without subsequent renormalization. A single antisymmetrized mean field determines the compound state.

Writing $\delta\rho_q=\rho_q-\rho_{q,\mathrm{ref}}$, the local normal-density constraint is
\begin{equation}
 \begin{split}
 E_{\rm var}={}&\int\dd^3x\,\varepsilon_{\rm SHF+BCS}\\
 &+\sum_{q=n,p}\int\dd^3x\,m
 \left[\lambda_q\delta\rho_q+\frac{C_q}{2}(\delta\rho_q)^2\right].
 \end{split}
 \label{eq:constraint}
\end{equation}
The support mask $m$ is one where the reference total density exceeds $10^{-10}$ fm$^{-3}$. The local multipliers $\lambda_q$ and stiffnesses $C_q=1000$ MeV fm$^3$ generate density-restoring fields. The density-constrained mean-field principle \cite{Umar2006} is used here with a static prescribed reference, following the construction of cluster-like intrinsic configurations in Ref.~\cite{Umar2023prc}. Reported path energies exclude the constraint terms. Center-of-mass corrections follow each functional's implemented prescription \cite{Bender2000CM}; the full correction is evaluated after variation. The multiplier ramp, mixing, and functional conventions are specified in Supplemental Material \cite{SupplementalMaterial}, Sec.~S1.

The surface-peaked contact pairing field is
\begin{equation}
 \Delta_q(\bm x)=-V_q\left(1-\frac{\rho_n+\rho_p}{\rho_0}\right)
 \chi_q^{\rm field}(\bm x), \label{eq:field}
\end{equation}
with positive magnitudes $V_n=999$, $V_p=1146$ MeV fm$^3$, and $\rho_0=0.16$ fm$^{-3}$. Zhou \textit{et al.}~\cite{Zhou2007} used these negative interaction strengths for medium-mass and heavy nuclei; here they are written as positive magnitudes. They also appear in the SkI4+BCS fit to five-point mass differences in Ref.~\cite{Bender2000Pairing}, Table~1. We retain them without refitting to SLy5 or the constrained $\Ti$ path. Mild strength variations are examined separately from particle--hole EDF dependence. The field profile depends on the anomalous source $\chi_q^{\rm field}$ as well as the density-dependent coupling. The smooth cutoff is
\begin{equation}
 f_k=\left[1+\exp\frac{\epsilon_k-\mu_q-E_{c,q}}{s_q}\right]^{-1}.
 \label{eq:cutoff}
\end{equation}
The windows are fixed once from the baseline 6-fm anchor: $(E_{c,n},s_n)\simeq(7.49,0.749)$ MeV and $(E_{c,p},s_p)\simeq(7.27,0.727)$ MeV. Their offsets from the channel chemical potential $\mu_q$ remain fixed in path and box comparisons. Couplings are not refitted for the other particle--hole functionals.

A squared anomalous-source lower bound $b_q=10^{-6}$ enters the baseline field evaluation through $\sqrt{\max[v_k^2(1-v_k^2),b_q]}$. The analyzed nonlocal tensor uses the solved $u_kv_k=\sqrt{v_k^2(1-v_k^2)}$, without this bound. Weak pairing requires checking the dependence of the solved response on the field-source bound. This numerical bound is separate from the energy cutoff and ultraviolet regularization of contact pairing. Supplemental Material, Secs.~S1, S5, and S8, specify its implementation and self-consistent controls.

The baseline spacing is $h=\Delta z=\Delta\rho=0.20$ fm, with $|z|\le15$ fm and cylindrical radius $\rho\le15$ fm. The density stopping measure is
\begin{equation}
 \eta_q=\left[\frac{\sum_i\omega_i m_i(\delta\rho_{q,i})^2}
 {\sum_i\omega_i m_i\rho_{q,\mathrm{ref},i}^2}\right]^{1/2}\le0.01,
 \label{eq:densityerror}
\end{equation}
where $\omega_i$ includes the cylindrical measure. The constraint-field mismatch must be below 0.001 MeV, the relative energy change below $10^{-6}$, and the occupied-state-weighted Hamiltonian residual below 0.05 MeV. Particle-number and boundary checks are also required. Strict controls at 6 and 7 fm use $\eta_q\le0.003$ and a field tolerance of 0.0003 MeV, with all other acceptance criteria unchanged. These criteria constrain relative $L^2$ density and iterative errors, not pointwise density equality.

Several Skyrme parametrizations are used to examine the model dependence. The SLy5, SkM*, and UNEDF1 calculations retain their implemented center-of-mass and spin-current prescriptions \cite{Bartel1982,Kortelainen2012}; SkI3, SkI4, and SkO are also evaluated with their original conventions \cite{ReinhardFlocard1995,Reinhard1999}. To isolate the role of the central quadratic spin-current sector, a matched SkI4 calculation restores the $J^2$ terms while keeping the original center-of-mass prescription. This variant is denoted SkI4+$J^2$. The restored terms originate from the central Skyrme interaction and are distinct from an additional tensor force \cite{Lesinski2007}. The reference density and pairing prescription are kept the same at each $d$. These comparisons probe the response to a prescribed geometry, rather than independently optimized cluster paths or separately refitted pairing functionals. The conventions and available calculations are summarized in Supplemental Material, Sec.~S1.

\subsection{Pairing tensor and distance-conditioned spatial distributions}
For real axial spinor amplitudes $U_k(\rho,z)$ and $D_k(\rho,z)$, the time-even singlet point kernel is
\begin{equation}
 \begin{split}
 \kappa_{s,q}(1,2)=\sum_{k>0} f_k u_kv_k\big[&U_k(1)U_k(2)\cos(m_k\varphi)\\
 &+D_k(1)D_k(2)\cos((m_k+1)\varphi)\big],
 \end{split} \label{eq:kernel}
\end{equation}
where $\varphi$ is the azimuthal difference and $m_k=\Omega_k-1/2$. The components are nonrelativistic spin amplitudes. Each time-reversed pair enters once through $k>0$, whereas the normal density counts $2v_k^2$. Equation~(\ref{eq:kernel}) defines a cutoff-regulated tensor in the same finite pairing space as the BCS calculation. A normalized spin-singlet coefficient $\sqrt{2}\kappa_{s,q}$ doubles its squared norm without changing normalized measures. Its scalar singlet norm is distinguished from the trace of the full spin-dependent anomalous density in Supplemental Material, Sec.~S8 B. The full tensor transforms covariantly when orbital phases and anomalous coefficients are transformed together.

Define the relative and center coordinates by
\begin{equation}
 \begin{gathered}
 \bm r=\bm r_1-\bm r_2,\qquad
 \bm R=\frac{\bm r_1+\bm r_2}{2}-z_0\hat{\bm z},\\
 r=|\bm r|,\quad R=|\bm R|,\quad Z=\bm R\cdot\hat{\bm z},
 \end{gathered} \label{eq:coordinates}
\end{equation}
where $z_0$ is the final total-nucleon-density center of mass. For each channel we compare
\begin{equation}
 X_{\kappa,q}=|\kappa_{s,q}|^2,\qquad
 X_{g,q}=\rho_q(\bm r_1)\rho_q(\bm r_2). \label{eq:weights}
\end{equation}
The latter samples two positions independently from the same normal density. Its full-support integral is $N_q^2$, with $N_n=24$ and $N_p=22$. It supplies a local-density geometry reference; an antisymmetrized normal two-body term would also require the off-diagonal one-body density. Exchange is deliberately absent, so subtraction of this reference does not isolate all pairing-induced correlations.

To characterize the spatial distributions within a relative-distance interval $B$, we define
\begin{align}
 I_X^B&=\int\dd^3r_1\dd^3r_2\,\bm1_B(r)X,\qquad
 f_X^B=I_X^B/I_X^{\rm all},\nonumber\\
 a_X^B&=\frac{\int\dd^3r_1\dd^3r_2\,\bm1_B(r)w_h(Z)X}{I_X^B},
 \label{eq:imbalance}\\
 \Delta a_q^B&=a_{\kappa,q}^B-a_{g,q}^B,\qquad
 C_{SL,q}=\Delta a_q^L-\Delta a_q^S. \label{eq:contrast}
\end{align}
The dimensionless center imbalance $a_X^B$ characterizes a distribution restricted to the distance interval $B$ and normalized within that interval. This is the meaning of distance-conditioned below. The intervals $S=[0,3)$, $M=[3,6)$, and $L=[6,\infty)$ fm define compact, intermediate, and extended components, respectively, with finite outer numerical support. Here, compact and extended are operational labels for relative-distance sectors of the same pairing tensor; the chosen boundaries serve as analysis cuts.
The allowed ranges are $a_X^B\in[-1,1]$, $\Delta a^B\in[-2,2]$, and $C_{SL}\in[-4,4]$; these differences are not percentages. The axial-cell classifier
\begin{equation}
 w_h(Z)=\operatorname{sgn}(Z)\left\{1-
 \left[\max\left(0,1-\frac{2|Z|}{h}\right)\right]^2\right\}
 \label{eq:classifier}
\end{equation}
assigns the central sampling layer symmetrically through a cell reconstruction. The point kernel and relative distance remain evaluated at the original nodes. A negative $\Delta a^B$ indicates a negative-end shift \emph{relative to the same-window reference}. The raw imbalance can still be positive: $a_\kappa^B<0$ is not required.

To connect with radial analyses \cite{Pillet2007,Pillet2010}, define the angular average $A_q(R,r)$ of $X_{\kappa,q}$ and
\begin{align}
 p_q(R,r)&=\frac{(4\pi)^2R^2r^2A_q(R,r)}{I_{\kappa,q}^{\rm all}},\nonumber\\
 r_{{\rm rms},q}^2&=\frac{\int r^2X_{\kappa,q}\,\dd^3r_1\dd^3r_2}{I_{\kappa,q}^{\rm all}}.
 \label{eq:radial}
\end{align}
The global $r_{\rm rms}$ is distinct from the local pair size $\xi(R)$. A joint signed-center density is
\begin{equation}
 \begin{split}
 F_X(Z,r)=\frac{1}{I_X^{\rm all}}\int\dd^3r_1\dd^3r_2\,X\,
 \delta(Z-R_z)\delta(r-|\bm r_1-\bm r_2|),\\
 H_q=F_{\kappa,q}-F_{g,q}.
 \end{split} \label{eq:joint}
\end{equation}
Both $p$ and $F$ have units fm$^{-2}$. Their normalization is over full support, even when a displayed panel is cropped. The joint difference $H$ contains both distance composition and center allocation; the window-normalized $\Delta a^B$ isolates the latter comparison.

All projected masses are deposited conservatively from the original integration weights:
\begin{equation}
 \begin{gathered}
 M_b=\sum_s C_{bs}W_sX_s,\qquad \sum_b C_{bs}=1,\\
 \sum_b M_b=\sum_s W_sX_s.
 \end{gathered}
 \label{eq:conservation}
\end{equation}
Overflow is retained in the integrals. Finite-volume displays use the cell masses directly, avoiding a second application of a node Jacobian after projection. Radial cells have width 0.20 fm; signed-center displays use $\Delta Z=0.40$ fm and $\Delta r=0.20$ fm. Azimuthal integration is analytic for the squared finite harmonic expansion and exact within the stored representation. Mesh checks separately assess spatial discretization. Supplemental Material, Sec.~S6, gives the projection and composition identities.

\subsection{Spatial partitions and microscopic diagnostics}
We partition the two nucleon coordinates into the alpha-side region $A$ and core-side region $C$, with a divider near the density neck. Fractions $f_{X,AA}^B$, $f_{X,AC}^B$ (including both orders), and $f_{X,CC}^B$ sum to one within each distance window. Their differences $\Delta f_T^B=f_{\kappa,T}^B-f_{g,T}^B$ locate anomalous excess or deficit in the compound state. The fractions describe endpoint positions rather than fragment occupations. Divider shifts of $\pm0.4$ fm test the partition sensitivity.

In the stored canonical basis, Eq.~(\ref{eq:kernel}) can be written as $\kap=\sum_k\kappa_{s,k}$. The incoherent sum, also called the self-only diagnostic, is defined by
\begin{equation}
 \begin{gathered}
 X_{\rm inc}=\sum_k|\kappa_{s,k}|^2,\\
 X_\kappa=X_{\rm inc}+2\operatorname{Re}\sum_{k<l}\kappa_{s,k}\kappa_{s,l}^*.
 \end{gathered}
 \label{eq:coherence}
\end{equation}
The incoherent sum omits interference between different canonical-pair amplitudes. Each distribution is normalized by its own intensity. This decomposition characterizes interference in the specified representation, whereas the full tensor obeys paired-basis covariance \cite{Ebata2010}. The phase, time-reversal, and allowed degenerate-subspace transformations are examined in Supplemental Material, Sec.~S3.

Overlap-matched SkI4 orbitals separate two sectors of a frozen linearized pairing kernel:
\begin{equation}
 {\cal K}_{ij}=\sqrt{\chi_i}\,G_{ij}\sqrt{\chi_j},\qquad
 \chi_i=\frac{f_i^2}{2|\epsilon_i-\mu_p|}. \label{eq:frozen}
\end{equation}
Here $G$ is the saved spatial pairing matrix. The $\chi$ sector uses the saved-state spectrum, saved chemical potential, and saved cutoff weights. Four exchanges combine original ($O$) and restored-$J^2$ ($J$) sectors in a common, overlap-qualified subspace. They diagnose frozen inputs without reoptimizing the density or solving a normal-state particle-number equation. Their construction and accuracy checks are given in Supplemental Material, Sec.~S4.

\section{Results and discussion}\label{Results}
\subsection{Geometry and energy of the controlled path}
\begin{figure*}[t]
 \includegraphics[scale=0.9]{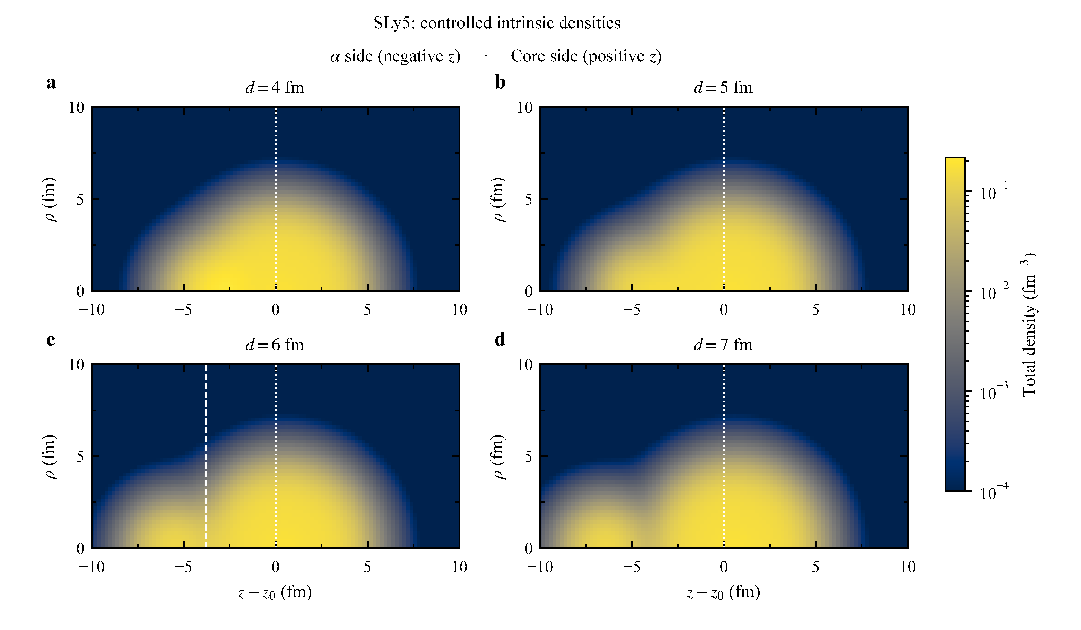}
 \caption{SLy5 total nucleon densities along the prescribed $\alpha+\Ca$-like intrinsic path at reference center separations $d=4$, 5, 6, and 7 fm. Each panel covers $-10\le z-z_0\le10$ fm and $0\le\rho\le10$ fm with equal coordinate scales. A common logarithmic color scale, with lower display limit $10^{-4}$ fm$^{-3}$, resolves the low-density connection. Dotted lines mark the final total-nucleon center; the 6-fm dashed line marks the endpoint divider. The alpha-side reference lies at negative $z$; the core side lies at positive $z$. Labels refer to reference centers, not surfaces. Reference and final axial profiles and residuals are compared in Supplemental Material, Sec.~S8.}
 \label{fig:geometry}
\end{figure*}

Figure~\ref{fig:geometry} displays the SLy5 normal densities that define the spatial setting. Increasing the reference center separation develops an unequal two-center environment and reduces the connection between its lobes. The alpha side remains on the negative axis. At small separation the translated reference components overlap strongly, so their labels do not imply independently identifiable fragments in the optimized density. We focus on 4--7 fm, where the path resolves the development of the asymmetric environment relevant to conditional pairing.

\begin{table}[t]
\caption{SLy5 energy context of the constrained path, in MeV. $E_{\rm sym}=-391.1752$ MeV is the lowest accepted unconstrained reflection-symmetric seed found here; $E_\alpha+E_{\rm Ca}=-383.0193$ MeV uses separately computed fragments with the same recipe. Constraint terms are excluded from $E(d)$.}
\label{tab:energy}
\begin{ruledtabular}
\begin{tabular}{crrr}
$d$ (fm)&$E(d)$&$E(d)-E_{\rm sym}$&$E(d)-E_\alpha-E_{\rm Ca}$\\
4&$-384.7138$&6.4614&$-1.6946$\\
5&$-383.5781$&7.5971&$-0.5588$\\
6&$-376.4387$&14.7365&6.5805\\
7&$-370.0564$&21.1188&12.9629
\end{tabular}
\end{ruledtabular}
\end{table}

The path energy increases with separation (Table~\ref{tab:energy}). At 6 and 7 fm it exceeds the separately computed fragment sum in this prescription. The unconstrained reference is the lowest accepted reflection-symmetric seed; an unrestricted ground-state search remains outside this calculation. The path supplies a controlled sequence for comparing intrinsic pairing distributions as the reference centers separate. Its members are constrained configurations, not established stationary cluster states. Assigning collective barriers, resonance energies, or laboratory excitation spectra would require dynamics and symmetry restoration.

\subsection{Distance-dependent spatial asymmetry of pairing correlations}
Integration over the relative distance conceals opposing neutron spatial preferences. At $d=6$ fm, the distance-integrated difference from the normal-density product is $\Delta a_n^{\rm all}=0.00995$, compared with a compact--extended contrast of $C_{SL,n}=0.4386$.

\begin{figure*}[t]
 \includegraphics[scale=0.9]{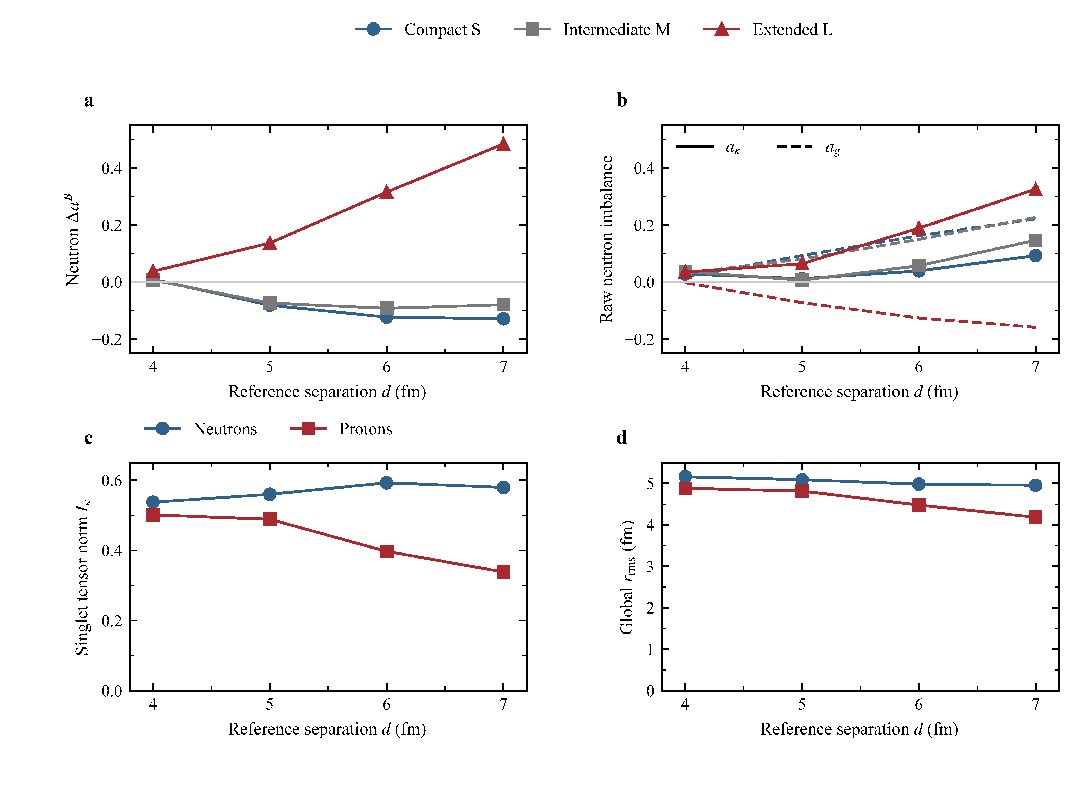}
 \caption{SLy5 path dependence. (a) Neutron center-imbalance differences for compact ($S$), intermediate ($M$), and extended ($L$) relative distances. (b) Raw neutron imbalances of the tensor (solid) and normal-density product (dashed), each normalized within the same interval. (c) Full scalar-singlet tensor norms and (d) global relative-distance rms values for both channels. Lines connect calculated configurations. The raw allocations distinguish the reference-relative sign in (a) from an absolute end preference. Proton imbalances and average gaps are given in the Supplemental Material.}
 \label{fig:path}
\end{figure*}

Figure~\ref{fig:path} shows how these quantities depend on the reference center separation along the SLy5 path, using identical spatial definitions for all configurations. The neutron compact and extended differences are negative and positive, respectively, at the sampled 5-, 6-, and 7-fm configurations, with
\[
 C_{SL,n}(d=5,6,7~{\rm fm})=0.2181,\;0.4386,\;0.6134.
\]
The compact difference indicates an alpha-side shift relative to the same-window reference; the extended difference indicates a core-side shift. From 6 to 7 fm, the extended difference increases by 0.169, while the compact difference changes by $-0.0057$. The contrast growth therefore comes chiefly from the extended reference-relative response. The intermediate component varies nonmonotonically.

The directional contrast grows as the global relative extent decreases. From 5 to 7 fm, $r_{{\rm rms},n}$ falls from 5.09 to 4.96 fm, while the extended fraction decreases from 0.278 to 0.262. Its anomalous weight peaks at 6 fm and then decreases by 3.21\%. In contrast, the bare-$uv$-weighted average neutron gap increases from 1.70 to 1.83 MeV over 5--7 fm; the occupation-weighted average also increases overall [Supplemental Material, Sec.~S8]. A larger gap, a growing conditional contrast, and a decreasing extended weight therefore coexist. The global rms describes the relative-distance distribution of the tensor and need not grow with the reference fragment separation.

The proton norm decreases from 0.501 at 4 fm to 0.338 at 7 fm, and its global relative-distance rms falls from 4.88 to 4.18 fm. Both compact and extended reference differences are negative at 5 fm; the extended difference becomes positive at 6 and 7 fm. The proton contrast reaches 0.472 at 7 fm. These channel differences concern conditional allocation as well as the overall anomalous weight. The much weaker SkI4+$J^2$ proton response and its source-bound dependence are examined in Sec.~III\,D.

\begin{figure*}[t]
 \includegraphics[width=\textwidth]{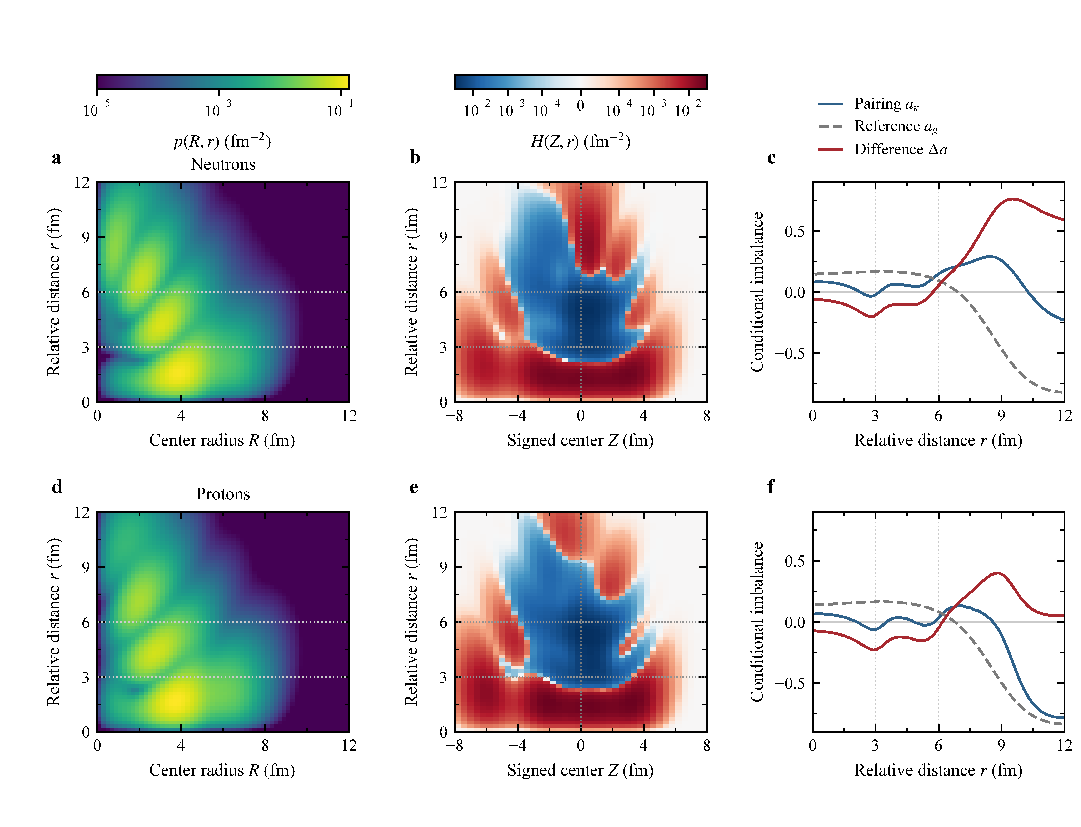}
 \caption{Full-support normalized spatial distributions of the SLy5 6-fm tensors: neutrons (upper row) and protons (lower row). Left: radial joint distribution $p(R,r)$ with a logarithmic color range $10^{-5}$--0.13569 fm$^{-2}$; zero radial mass is white. Center: $H(Z,r)=F_\kappa-F_g$, with a symmetric logarithmic color range $\pm0.033585$ fm$^{-2}$ that is linear for $|H|<10^{-4}$ fm$^{-2}$. These color scales resolve low-weight tails and weak signed structure; all spatial coordinate axes remain linear. $H$ includes distance composition and center allocation, so colors do not directly quantify $\Delta a^B$. Right: distance-bin imbalances $a_\kappa(r)$, $a_g(r)$, and their difference. Dotted lines locate the 3- and 6-fm distance cuts. All displayed distance bins have support; unsupported bins would remain blank. Cropped maps retain the full-support normalization.}
 \label{fig:joint}
\end{figure*}

Figure~\ref{fig:joint} compares radial and signed-center distributions at 6 fm. Radial panels locate the relative-distance support after averaging over the center direction. The joint difference $H(Z,r)$ retains excess and deficit at the two ends, together with changes in distance composition. The conditional curves separate the raw tensor imbalance from the normal-density-reference imbalance. Their difference measures a relative spatial preference within each distance interval.

For compact neutrons at 6 fm, $a_\kappa^S=0.0390$ and $a_g^S=0.1623$ give $\Delta a_n^S=-0.1233$. The negative difference is a relative alpha-side shift, although the raw tensor imbalance is positive. Extended neutrons have $a_\kappa^L=0.1888$ and $a_g^L=-0.1264$, yielding $\Delta a_n^L=0.3152$. Extended protons illustrate the converse case: $a_\kappa^L=-0.0530$ and $a_g^L=-0.1506$ give a positive difference, $\Delta a_p^L=0.0975$, between two negative imbalances.

From 6 to 7 fm, the extended anomalous imbalance increases by 0.137 and the reference imbalance decreases by 0.0324. Both changes enlarge $\Delta a_n^L$. The compact raw imbalances instead increase together, by 0.0535 and 0.0591, leaving their difference nearly stable. Defining $C_{\kappa,SL}=a_\kappa^L-a_\kappa^S$ and $C_{g,SL}=a_g^L-a_g^S$ gives $C_{SL}=C_{\kappa,SL}-C_{g,SL}$. Their contributions to its growth are $\delta C_{\kappa,SL}=+0.0833$ and $-\delta C_{g,SL}=+0.0916$. The anomalous and reference changes are both substantial [Supplemental Material, Sec.~S6].

Strict-density states preserve this response. Their neutron contrast grows from 0.4433 to 0.6107 between 6 and 7 fm, with opposing extended raw changes and similar compact raw changes. The contrast increment is 4.26\% smaller than at baseline precision, while the extended anomalous weight decreases by 2.98\% [Supplemental Material, Sec.~S6 and Table~S6]. Tighter density matching therefore preserves the distinction between conditional redistribution and strength.

Distance integration combines window allocation with distance composition through $a_X^{\rm all}=\sum_B f_X^Ba_X^B$. For neutrons at 6 fm, the weighted compact, intermediate, and extended conditional differences nearly cancel, leaving 0.00218. Unequal distance fractions add 0.00777 to give 0.00995 [Supplemental Material, Sec.~S6, Eq.~(S5)]. The raw integrated imbalances remain finite: $a_{\kappa,n}^{\rm all}=0.08409$ and $a_{g,n}^{\rm all}=0.07414$. The small difference between them conceals the opposing conditional allocations.

\subsection{Spatial partitions and interference in the canonical basis}
\begin{figure*}[t]
 \includegraphics[scale=0.9]{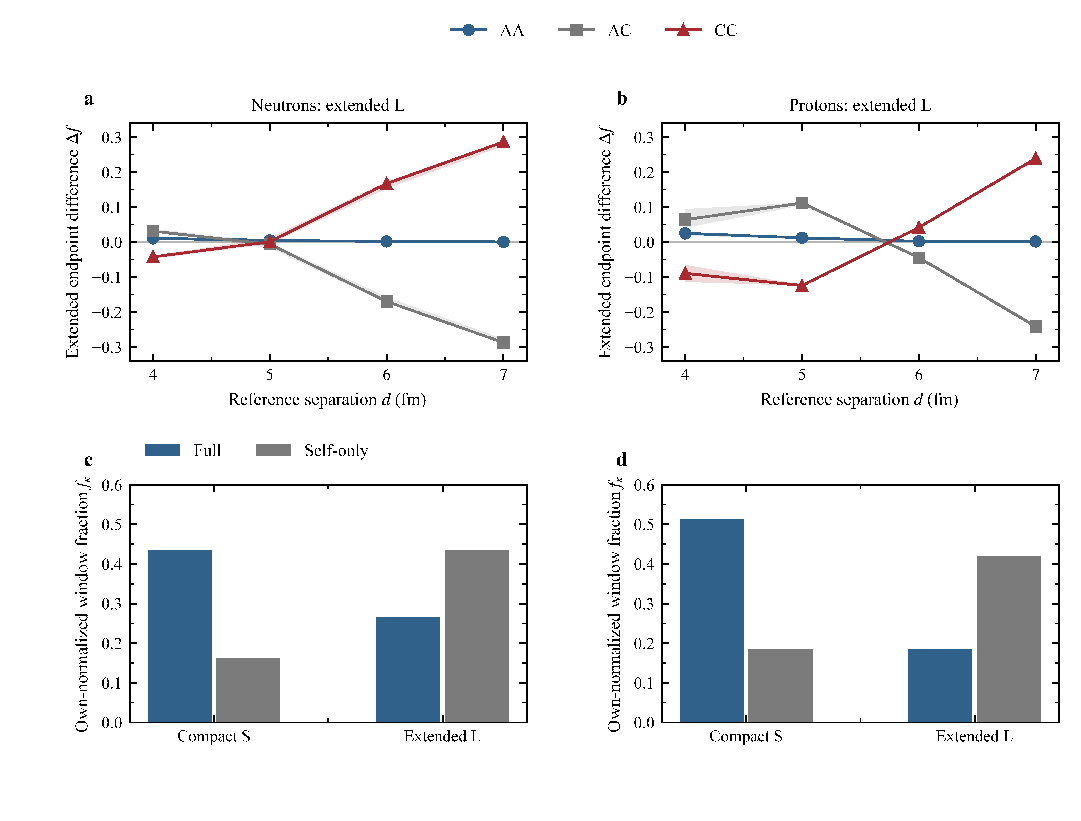}
 \caption{Spatial partitions and canonical-pair interference for SLy5. Upper panels: endpoint-fraction differences relative to the normal-density product for extended neutron and proton correlations along the path. $AA$, $AC$, and $CC$ denote two alpha-side, opposite-side, and two core-side endpoints. The divider variation tests the spatial partition. Lower panels: compact and extended fractions of the 6-fm full tensor and the representation-dependent incoherent sum in Eq.~(\ref{eq:coherence}), each normalized by its own intensity. The two comparisons resolve nucleon positions and interference between canonical-pair amplitudes, respectively.}
 \label{fig:endpoint}
\end{figure*}

The pair-center coordinate alone does not determine where the two nucleons are located. A large relative distance can connect positions on opposite sides or two widely separated positions within the larger core-side region. To distinguish these possibilities, Fig.~\ref{fig:endpoint} shows the endpoint fractions. For extended neutrons, the core-side excess $\Delta f_{CC}^L$ reaches 0.1676 at 6 fm and 0.2866 at 7 fm. The anomalous fraction increases from 0.7922 to 0.8853, whereas the reference fraction decreases from 0.6246 to 0.5987. Shifting the divider by $\pm0.4$ fm gives excess ranges of 0.1536--0.1733 and 0.2755--0.2901, respectively. Both the baseline and strict-density calculations therefore support an excess within the core-side region at these separations [Supplemental Material, Table~S6]. At 5 fm, the range includes zero and the sign is unresolved. Since the divider follows the neck of each configuration, these fractions describe evolving spatial partitions rather than transport through a fixed plane.

Extended protons also have a positive $CC$ excess at 6 fm, with a smaller magnitude of 0.0419. Compact correlations have a core-side deficit in both channels. The neutron response thus includes excess extended weight with both endpoints on the core side. The $AA$ extended fraction is small but nonzero, about $2.45\times10^{-3}$ at 6 fm for neutrons. A narrow axial alpha-side region does not exclude large three-dimensional separations, which also depend on transverse positions and azimuthal angles. These partitions locate compound-state correlations rather than isolated-fragment occupations.

The lower panels of Fig.~\ref{fig:endpoint} examine the role of interference between canonical-pair amplitudes. At 6 fm, the neutron compact fraction increases from 0.1629 in the incoherent sum to 0.4357 in the full tensor, while the extended fraction decreases from 0.4364 to 0.2650. The corresponding proton changes are 0.1855 to 0.5141 and 0.4205 to 0.1857. Thus the coherent cross terms concentrate the normalized relative-distance distribution in the compact component for both channels in the stored basis. This behavior is consistent with the many-orbital origin of compact correlations found in coordinate-space studies \cite{Matsuo2005}; here the decomposition concerns a particular constrained state and canonical representation.

The saved-canonical self-only neutron distribution retains opposite compact and extended preferences at all three 5--7-fm states, with contrast increasing from 0.2041 to 0.4782. Cross terms increase the full neutron contrast at each point. They reduce proton contrast at 5 and 6 fm and enhance it at 7 fm. Each distribution uses its own intensity for normalization [Supplemental Material, Sec.~S3, Table~S10].

Phase and time-reversal tests verify covariance of the full 6-fm tensor. No nontrivial same-$\Omega$ degenerate block meets the saved-state rotation tolerances. A synthetic equal-weight block preserves the full tensor under rotation but changes the self-only pattern [Supplemental Material, Sec.~S3]. The interference decomposition is consequently representation-specific; the primary spatial result follows from the full tensor.

\subsection{Model dependence and microscopic origin of the anomalous response}
\begin{figure*}[t]
 \includegraphics[scale=0.9]{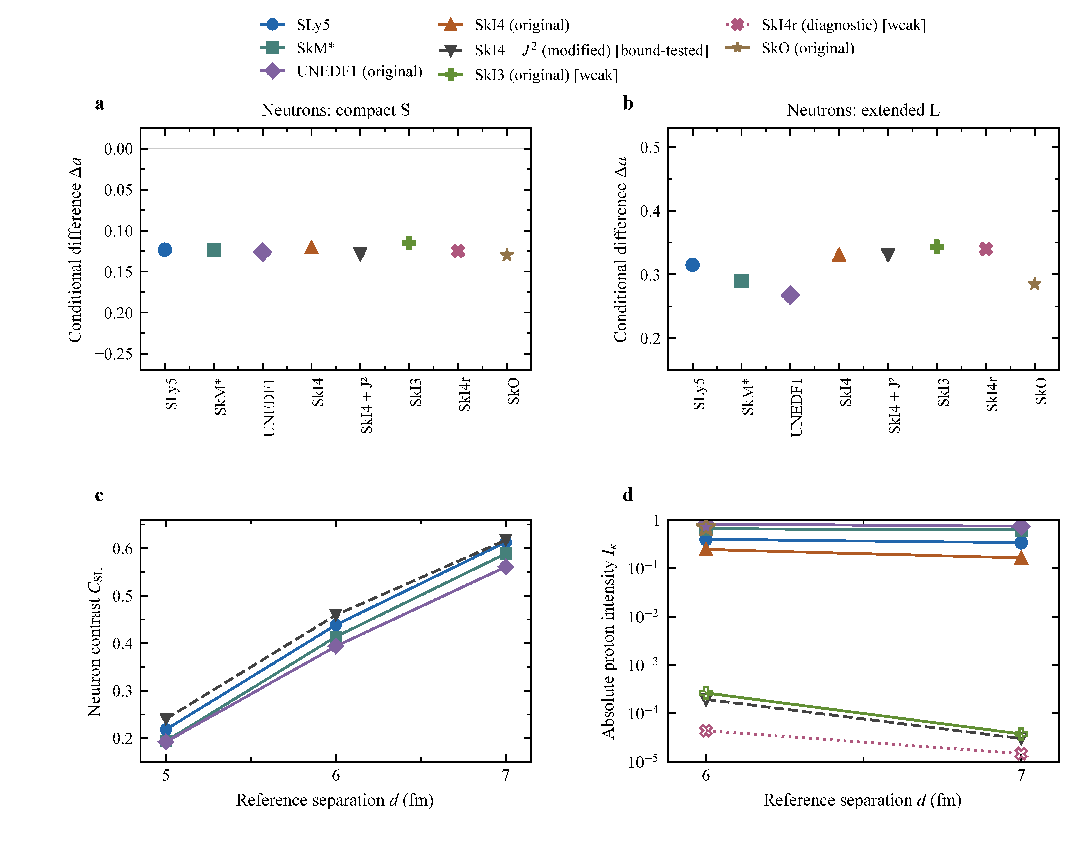}
 \caption{Model dependence under common pairing strengths and fixed channel windows, without EDF-specific pairing refits. Upper panels: compact and extended neutron differences at 6 fm. Lower left: neutron contrasts along the completed controlled 5--7-fm sequences. Lower right: proton norms on a logarithmic scale. Open weak-channel markers distinguish SkI4+$J^2$, SkI3, and SkI4r. The bound-tested label refers specifically to the SkI4+$J^2$ 7-fm control; individual source-bound tests are unavailable for SkI3 and SkI4r. SkO has one accepted 6-fm point; its rejected 7-fm case is omitted. Original-convention neutron trends appear in Supplemental Material, Sec.~S1.}
 \label{fig:models}
\end{figure*}

Figure~\ref{fig:models} compares the conditional distributions obtained with different microscopic prescriptions. The accepted original-convention and controlled calculations retain a negative compact and positive extended neutron difference at 6 and 7 fm. In the completed controlled sequences, the neutron contrast increases over the available separations. For original SkI3 and SkI4, it increases from 0.4591 and 0.4516 at 6 fm to approximately 0.6252 and 0.6212 at 7 fm, respectively. The distance-selective neutron pattern is therefore stable within the tested prescriptions, although its magnitude and the detailed distributions remain model dependent.

The density constraint does not fix the spin-current density, orbital occupations, or wave-function overlaps. A change of the central quadratic spin-current sector modifies the single-particle spin-orbit field and allows these quantities to readjust \cite{Reinhard1999,Lesinski2007}. Nearly matched normal-density distributions can therefore coexist with markedly different anomalous amplitudes. The neutron--proton comparison illustrates this microscopic freedom within the prescribed spatial environment.

Proton norms vary strongly when the common pairing strength is retained without EDF-specific refits. SLy5, SkM*, UNEDF1, original SkI4, and the accepted SkO state retain appreciable anomalous support. Original SkI4 gives $I_{\kappa,p}=0.245$ and 0.164 at 6 and 7 fm; restoring its central spin-current sector gives $1.95\times10^{-4}$ and $3.02\times10^{-5}$. SkI3 and the spin-orbit diagnostic SkI4r also have small amplitudes, with their source-bound dependence untested. The matched SkI4 comparison isolates a change of spin-current prescription at common pairing parameters; the broader EDF spread also includes differences between independently fitted particle--hole models.

\begin{figure*}[t]
 \includegraphics[scale=0.9]{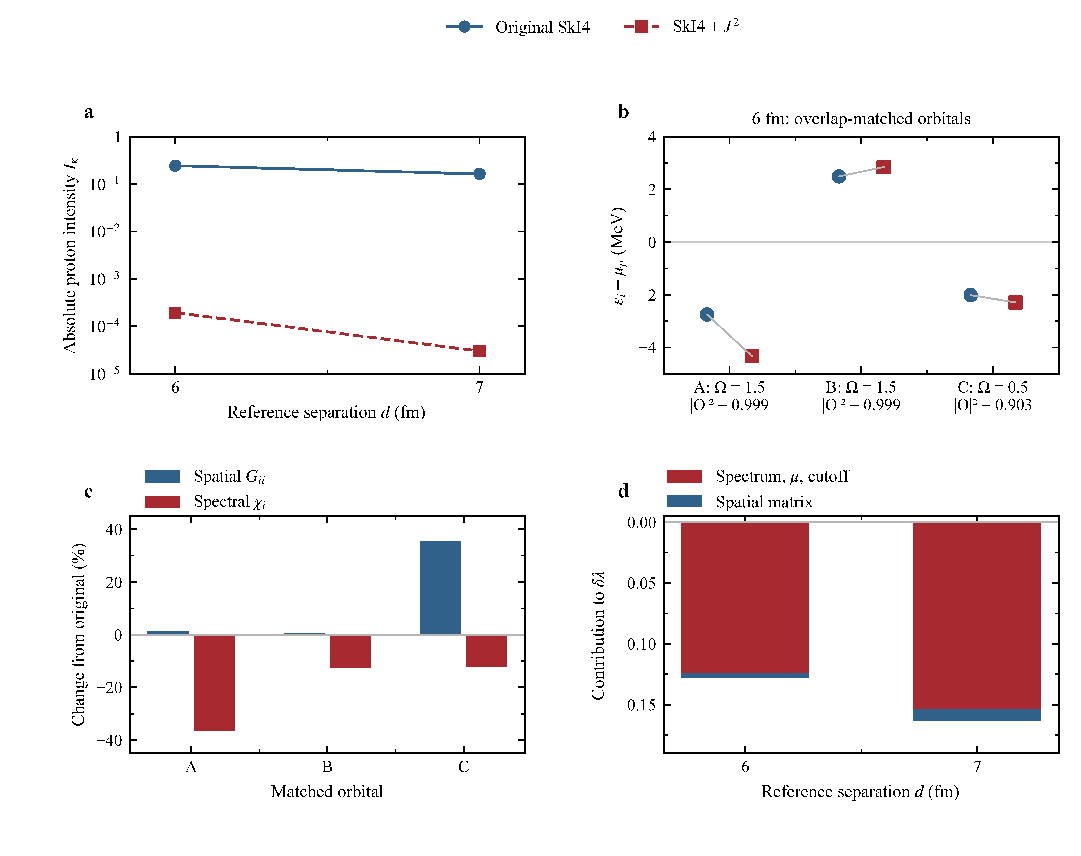}
 \caption{Microscopic analysis of the matched SkI4 proton calculations. (a) Scalar-singlet norms. (b) Overlap-matched energies relative to each saved chemical potential. (c) Percentage changes of diagonal spatial pairing matrices and spectral susceptibilities, including the larger spatial response of match C. (d) Two-order averaged contributions to the frozen-kernel eigenvalue change. $O$ and $J$ denote original SkI4 and SkI4+$J^2$. The four underlying common-subspace exchanges are given in Supplemental Material, Table~S11. This is a saved-input diagnostic without a new particle-number solution.}
 \label{fig:ski4}
\end{figure*}

The matched SkI4 comparison exposes how that support changes. Original and restored-$J^2$ states share the center-of-mass correction, pairing recipe, reference density, and numerical criteria. Figure~\ref{fig:ski4} tracks same-channel, same-$\Omega$ orbitals by spatial overlap. At 6 fm, match A has squared overlap 0.99946, while $\epsilon-\mu_p$ moves from $-2.75$ to $-4.33$ MeV. Its diagonal pairing matrix increases by 1.46\%, but its susceptibility decreases by 36.5\%. Match B has a 0.74\% matrix increase and a 12.6\% susceptibility decrease. Match C, with squared overlap 0.9032, instead has a substantial 35.6\% matrix increase and a 12.2\% susceptibility decrease. Spatial changes are consequently nonuniform across orbitals; the whole-subspace exchange tests their combined contribution.

The frozen exchanges quantify the response across the overlap-qualified orbital subspace. At 6 fm, the leading eigenvalues are $\lambda_{OO}=1.0998$, $\lambda_{OJ}=1.0947$, $\lambda_{JO}=0.9749$, and $\lambda_{JJ}=0.9721$. The first index specifies the spectrum, saved chemical potential, and cutoff sector, whose two-order averaged contribution is
\[
 \delta\lambda_\chi=
 \frac{(\lambda_{JO}-\lambda_{OO})+(\lambda_{JJ}-\lambda_{OJ})}{2}.
\]
The ratio $\delta\lambda_\chi/(\lambda_{JJ}-\lambda_{OO})$ is 96.9\% at 6 fm and 94.4\% at 7 fm. Its denominator is the auxiliary eigenvalue change, distinct from the nonlinear tensor-norm suppression. The spatial-matrix sector supplies the complementary contribution. The mixed finite difference is about 1.8\% and 2.1\% of the total eigenvalue change, respectively, and is shared by the two-order average [Supplemental Material, Sec.~S4, Table~S11]. This budget assigns the dominant change to the spectrum, chemical potential, and cutoff sector while retaining the orbital-dependent spatial response.

Relative truncation errors in the proton own-input leading eigenvalues are below $1.6\times10^{-8}$. The saved gap contraction reproduces printed gaps within their 0.005-MeV rounding resolution. The analogous 6-fm neutron exchange is excluded because it fails the common-subspace quality criterion. The full-space restored-$J^2$ proton kernel is sensitive to the frozen chemical potential. Three alternative placements at 6 fm give eigenvalues 1.156, 0.980, and 1.273 with saved cutoff weights fixed [Supplemental Material, Sec.~S4]. These unity crossings characterize saved-input susceptibility. Establishing a self-consistent phase transition would also require solving the normal-state particle-number equation.

Self-consistent controls identify the residual SkI4+$J^2$ response at 7 fm [Supplemental Material, Sec.~S5, Table~S12]. Tightening stopping alone changes proton $C_{SL}$ from 0.4906 to 0.4088, a shift of $-0.0818$. Under tight stopping, lowering the squared source bound from $10^{-6}$ to $10^{-8}$ reduces $I_{\kappa,p}$ by a factor of 99.72 (99.00\%). The 15-fm contrast changes by 0.00035. Enlarging the low-bound box to 18 fm changes it by 0.0804, failing the weak-proton shape criteria; neutrons pass the matched control. The intensity scaling identifies a source-bound-maintained near-normal response. Its normalized shape remains box sensitive, precluding a resolved finite-condensate interpretation. This source-bound control covers SkI4+$J^2$, with SkI3 and SkI4r limits still undetermined.

Surface-pairing SLy5 retains appreciable anomalous support under the tested source and orbital-space controls. Lowering $b_p$ from $10^{-6}$ to $10^{-8}$ at 6 and 7 fm, with $b_n$ unchanged, changes proton norms by less than 0.04\% and neutron norms by less than 0.008\%. In both channels, the same-window differences change by less than $1.8\times10^{-4}$ and $C_{SL}$ by less than $6.4\times10^{-5}$. Expanding the 6-fm orbital pool from 41 proton and 44 neutron time-reversed pairs to 65 and 67, respectively, changes the same-window differences by less than $8\times10^{-6}$ [Supplemental Material, Sec.~S8, Tables~S17 and S18]. The conditional SLy5 pattern thus persists under these controls, while the weak SkI4+$J^2$ proton amplitude scales with the field-source bound.

Matched SLy5 6-fm box, mesh, and mirror controls support the integrated spatial measures. Tighter density matching at 6 and 7 fm preserves the neutron conditional response and reduces proton norms by about 4.3--4.4\%. Distance-cut and divider variations preserve the 6--7-fm neutron interpretation. The deterministic sensitivity budget and local-map comparison are given in Supplemental Material, Secs.~S2 and S6. These representative SLy5 tests do not constitute independent mesh and box controls for every EDF, including SkO.

\subsection{Orbital support of the neutron spatial response}
The conditional neutron response can be resolved further within the stored canonical representation. We label orbitals by $K\equiv\Omega$ and track them between the saved 6- and 7-fm states by their spinor overlaps within the same $K$ block. To retain the interference with all other orbitals, define the signed allocation
\begin{equation}
 \mathcal D_{k,q}^{B}=
 \frac{\operatorname{Re}\int_B\dd^3r_1\dd^3r_2\,
 w_h(Z)\kappa_{s,k}\kap^*}{I_{\kappa,q}^{B}},
 \qquad \sum_k\mathcal D_{k,q}^{B}=a_{\kappa,q}^{B}.
 \label{eq:orbitalallocation}
\end{equation}
Each cross term is shared equally between its two participants. Contributions to $C_{\kappa,SL}$ are therefore $\mathcal D_k^L-\mathcal D_k^S$, with common full-tensor window denominators. These signed allocations retain interference and can be negative; orbital probabilities require a different definition. The normal-density-reference part of $C_{SL}$ remains separate.

\begin{figure*}[t]
 \includegraphics[scale=0.9]{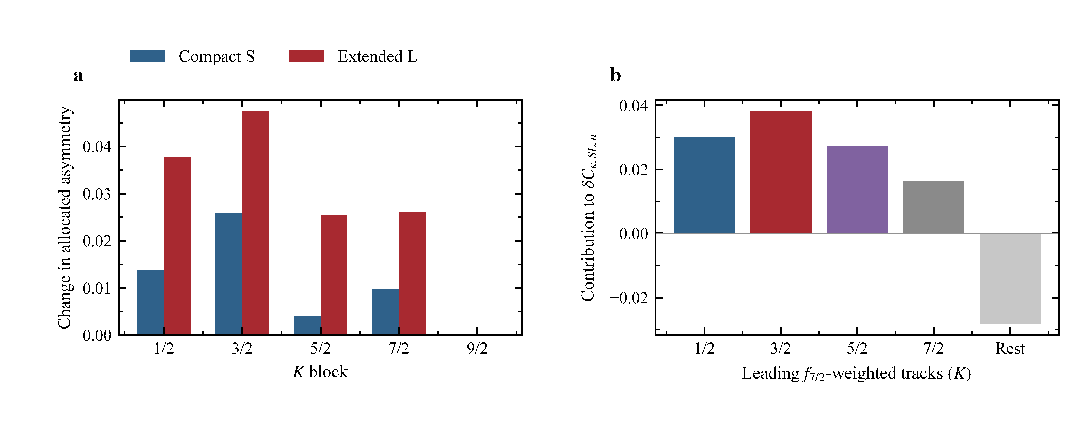}
 \caption{Orbital support of the SLy5 neutron response between 6 and 7 fm. (a) Changes of the compact and extended allocations in Eq.~(\ref{eq:orbitalallocation}), summed within every stored $K$ block. The $K=9/2$ contribution is retained although its magnitude is below $4\times10^{-9}$. (b) Contributions to $\delta C_{\kappa,SL,n}$ from the four leading positive tracks and from all remaining orbitals. Each signed contribution includes its shared interference with the full tensor and uses the same full-window denominators. The leading tracks have predominantly $f_{7/2}$ angular components. Their pairing coefficients and single-particle spatial properties are given in Supplemental Material, Table~S14 and Sec.~S8.}
 \label{fig:orbitalresponse}
\end{figure*}

Figure~\ref{fig:orbitalresponse} resolves the baseline 6-to-7-fm change into orbital contributions. In each appreciable $K=1/2,3/2,5/2,7/2$ block, the extended anomalous allocation increases more than the compact allocation. The four largest positive individual increments have predominantly $f_{7/2}$ angular components. Their weights are 66.5--96.5\% at 6 fm and 74.3--95.3\% at 7 fm; squared tracking overlaps range from 0.9403 to 0.9988. Contributions to $\delta C_{\kappa,SL,n}$ are 0.03001, 0.03833, 0.02714, and 0.01619, respectively. Their sum, 0.11166, combines with $-0.02835$ from the remaining orbitals to recover the anomalous increment 0.08331. Leading $f_{7/2}$-weighted branches therefore contribute alongside a counteracting orbital background.

All four tracks acquire greater core-side single-particle weight, with different pairing-coefficient responses. For $K=3/2$, the core-side weight rises from 0.8699 to 0.9534 as the stored coefficient decreases from 0.4843 to 0.4544. Its mean axial position relative to the compound density center changes from $-0.1488$ to $+0.3332$ fm, independently of the endpoint divider. The $K=5/2$ and $7/2$ coefficients change little, while their extended directional allocations increase. Spatial readjustment thus contributes to the response alongside amplitude changes. A unique separation of occupation and wave-function effects would require an additional controlled comparison. The angular components are defined about the compound density center and include opposite-parity admixtures. We use $f_{7/2}$-dominated labels with the radial quantum number left unassigned. Supplemental Material, Sec.~S7 and Table~S14, documents the 5-fm tracking check and numerical controls.

Angular analysis also identifies the matched SkI4 $K=3/2$ proton hole as predominantly $d_{3/2}$ and the nearby particle track as predominantly $f_{7/2}$ (Table~S14). Restoring the central $J^2$ terms preserves these dominant characters while reducing their saved-state spectral susceptibility. The labels specify the near-Fermi branches in Fig.~\ref{fig:ski4}; quantitative attribution remains with the qualified whole-subspace frozen-kernel budget. A related sensitivity to the single-particle environment appears in neutron-rich Ba, where tensor-induced changes in spin-orbit spacings modify preexisting octupole correlations \cite{Xue2026Ba}. The present comparison concerns the central spin-current terms and pairing support at constrained density, rather than an added tensor interaction or an optimized octupole minimum.

\section{Physical interpretation and scope}\label{Physical}
Relative distance and pair-center position are coupled in the intrinsic tensor. Integrating over distance mixes components with different reference-relative allocations, which nearly cancel at 6 fm. Between the 6- and 7-fm configurations, the extended anomalous imbalance increases while the normal-density-reference imbalance decreases, strengthening the core-side preference relative to the reference. Over the same interval in the surface-pairing baseline, the extended anomalous weight decreases and the average neutron gap increases. Much of its weight has both endpoints on the core side. This core-side geometry shows that large relative distances also occur between two positions on the same side of the neck. The orbital budget connects the response to spatial readjustments of $f_{7/2}$-weighted branches with different amplitude changes.

The raw anomalous imbalance and its difference from the normal-density product answer distinct questions. For example, compact neutrons at 6 fm have a small positive raw imbalance but a negative reference-relative difference. Both distributions evolve along the path, so their separate imbalances are needed to interpret the growth of the contrast. Subtracting the independent-coordinate reference retains effects of spectra, occupations, localization, and interference together with residual geometry. Resolving these allocations at fixed relative distance extends radial analyses of spatial pairing \cite{Pillet2007,Pillet2010} to unequal ends of a prescribed intrinsic configuration.

At common fixed cutoffs and prescribed reference densities, the volume-pairing comparison at 6 and 7 fm preserves the opposing reference-relative neutron shifts and their growing contrast [Supplemental Material, Sec.~S9]. Its extended anomalous weight rises over this interval, whereas the surface-pairing baseline weight falls. The weak volume-pairing proton channel remains a source-sensitive prescription response.

The neutron conditional pattern persists under the tested particle--hole and mild pairing variations. Proton support is more sensitive to the fixed microscopic prescription. The matched SkI4 budget assigns most of the auxiliary eigenvalue change to its spectral sector, including the saved chemical potential and cutoff. Self-consistent controls identify source-bound maintenance specifically for SkI4+$J^2$ at 7 fm. Cutoff dependence and continuum treatment limit extrapolation beyond the present finite-space BCS prescription \cite{BulgacYu2002,Borycki2006,Dobaczewski1996}.

The diagnostic concerns two-nucleon anomalous correlations in a cluster-like environment; four-nucleon correlations and alpha-condensate wave functions address a different many-body structure \cite{Tohsaki2001}. The signed asymmetries characterize reflection-asymmetric \emph{intrinsic} states and reverse under strict reflection of the reference and calculated state. The corresponding parity-odd expectation vanishes in a laboratory parity eigenstate. Connecting these diagnostics to observables requires symmetry restoration and transition or reaction calculations \cite{Sheikh2021}. Variation after parity projection followed by angular-momentum projection offers an example of connecting asymmetric Skyrme intrinsic configurations to spectra \cite{Ohta2004prc}. Angular-momentum projection and configuration mixing have also connected deformed mean-field configurations to the low-lying spectra of C and Ne nuclei \cite{Xue2024C12,Xue2024Ne}. The nonlocal tensor itself is an anomalous mean-field quantity, with a measured pair probability requiring a different construction. Within the prescribed path and fixed BCS treatment, its distance-conditioned structure resolves spatial information relevant to those future studies. Resonance assignments and transfer strengths remain questions for the corresponding dynamical and symmetry-restored calculations.

\section{Summary and outlook}\label{Summary}
Density-constrained SHF+BCS calculations along a prescribed $\alpha+\Ca$-like path in $\Ti$ reveal oppositely signed compact and extended neutron shifts relative to the same-window normal-density product. Their near cancellation at 6 fm leaves a small distance-integrated difference despite a substantial conditional contrast. From 6 to 7 fm, the extended anomalous imbalance increases while its normal-density reference imbalance decreases, enlarging their difference. Over the same interval in the surface-pairing baseline, the extended anomalous weight decreases and the average neutron gap increases. Distance-conditioned allocation therefore contains structural information absent from a gap, tensor norm, or global relative size.

Endpoint partitions show excess extended neutron weight with both endpoints on the core side at 6 and 7 fm. In the stored canonical representation, interference enhances compact support, and leading $f_{7/2}$-weighted branches acquire greater core-side probability with distinct amplitude responses. The self-only tensor already supports the opposing reference-relative signs; interference redistributes this pattern. The normal-reference change contributes separately to the full contrast growth.

The neutron pattern persists within the tested prescriptions, while proton norms depend strongly on microscopic pairing support. In matched SkI4 calculations, the spectrum, chemical potential, and cutoff dominate the auxiliary frozen-kernel eigenvalue response. The weak SkI4+$J^2$ proton configuration at 7 fm exhibits a source-bound-maintained near-normal response.

Within the prescribed intrinsic path, distance-conditioned distributions resolve how neutron pairing weight is allocated between unequal ends, beyond what gap and size measures convey. Endpoint partitions and interference-retaining orbital budgets identify the spatial and single-particle support of this organization.

The analysis provides a robust diagnostic to examine pairing in other intrinsic configurations with inequivalent spatial regions. In reflection-asymmetric nuclei, comparisons along isotopic chains and deformation paths could determine how shell filling and shape affect the relative-distance dependence of the pair-center distribution. For configurations with a developed neck, extending the endpoint partitions to the two fragments and the neck region would help identify where anomalous weight is concentrated. Furthermore, applying the analysis to coordinate-space HFB states with a controlled continuum treatment would extend these questions to weakly bound systems, and combining distance-resolved analysis with symmetry restoration would provide a route to connect intrinsic spatial correlations to laboratory observables.

\section*{Acknowledgements}
This work was supported by the National Natural Science Foundation of China under 
Grants No. 12505137, 12305144, and the Key Scientific Research Projects of Universities in Henan Province (Grant No. 26B140011).

\bibliography{neutron_pair_directionality}
\end{document}